\documentclass[pra,twocolumn,showpacs,preprintnumbers,amsmath,amssymb]{revtex4}
\makeatletter
\usepackage[dvips]{graphicx}	

\usepackage{amsmath}
\usepackage{amssymb}
\usepackage{amsbsy}
\usepackage{color}
\setcitestyle{square}
\usepackage{epstopdf}
\DeclareGraphicsExtensions{.pdf,.eps,.png,.jpg,.mps}

\begin{document}

\title{
Single layer graphene with electronic properties as gated bilayer}
\author{W. Jask\'olski}
\email{wj@fizyka.umk.pl}
\affiliation{Institute of Physics, Faculty of Physics, Astronomy and Informatics, Nicolaus Copernicus University, Grudziadzka 5, 87-100 Toru\'n, Poland}

\date{\today}


\begin{abstract} 
We demonstrate that single layer graphene exhibits the electronic structure of a bilayer when it is connected to two gated bilayers. The energy gap characteristic for gated bilayer is induced in the single layer and it persists for distances separating the bilayers up to 20 nm. Most importantly, we report on the appearance of topologically protected gapless state localized in a single graphene layer when the distant bilayers differ in the stacking order. The one-dimensional current in the topological state flows in a single layer graphene, and this finding suggests a new exploitation of graphene in nanoelectronics. 
\end{abstract}

\pacs{73.63.-b, 72.80.Vp}

\maketitle

\section{\label{sec:intro} Introduction}

One of the obstacles in the common application of graphene in nanoelectronics is the gapless electronic structure of the single layer graphene (SLG). Various ways have been proposed to brake this bottleneck  and to force graphene to open the energy gap. One of them is to apply a gate voltage to the Bernal-stacked bilayer graphene (BLG) \cite{ Ohta_2006,Castro_2007,Oostinga_2008,Zhang_2009,Szafranek_2010}. The value of the gap in gated BLG is to some extent tunable with the gate voltage, which is important for electronic applications \cite{Padilha_2011,Schwierz_2010,Lin_2008,Choi_2010,Santos_2012,Zhang_transistor_2018} as well as for the study of fundamental phenomena in graphene  \cite{Maher_2013,Martin_2008,Barlas_2010,Zhang_2013,Vaezi_2013}. 

Bilayer graphene focuses recently a lot of attention do to its superconducting properties discovered for twisted layers \cite{Jarillo2018}. 
Gated bilayer has an additional advantage: it provides topologically protected gapless states when stacking domain walls (the change from AB to BA of the stacking order of sublattices) are present in the BLG \cite{Vaezi_2013, Alden_2013,San_Jose_2014,Pelc_2015}. The pair of states connecting the valence and conduction band continua appears at each valley ($K$ and $K^{\prime}$), they are localized at the domain walls and  provide one-dimensional topologically protected currents along the stacking boundary. The stacking domain walls in the BLG can be created by  stretching, corrugating or delaminating one of the layers \cite{Lin_2013,Ju_2015,Pelc_2015,Peeters_2018}. Gapless states persist in gated BLG even when the stacking domain walls contain atomistic defects \cite{Jaskolski_2016}. The existence of topologically protected gapless states has been already reported in experiments \cite{Ju_2015,Li_2016}. 

\begin{figure}[ht]
\centering
\includegraphics[width=\columnwidth]{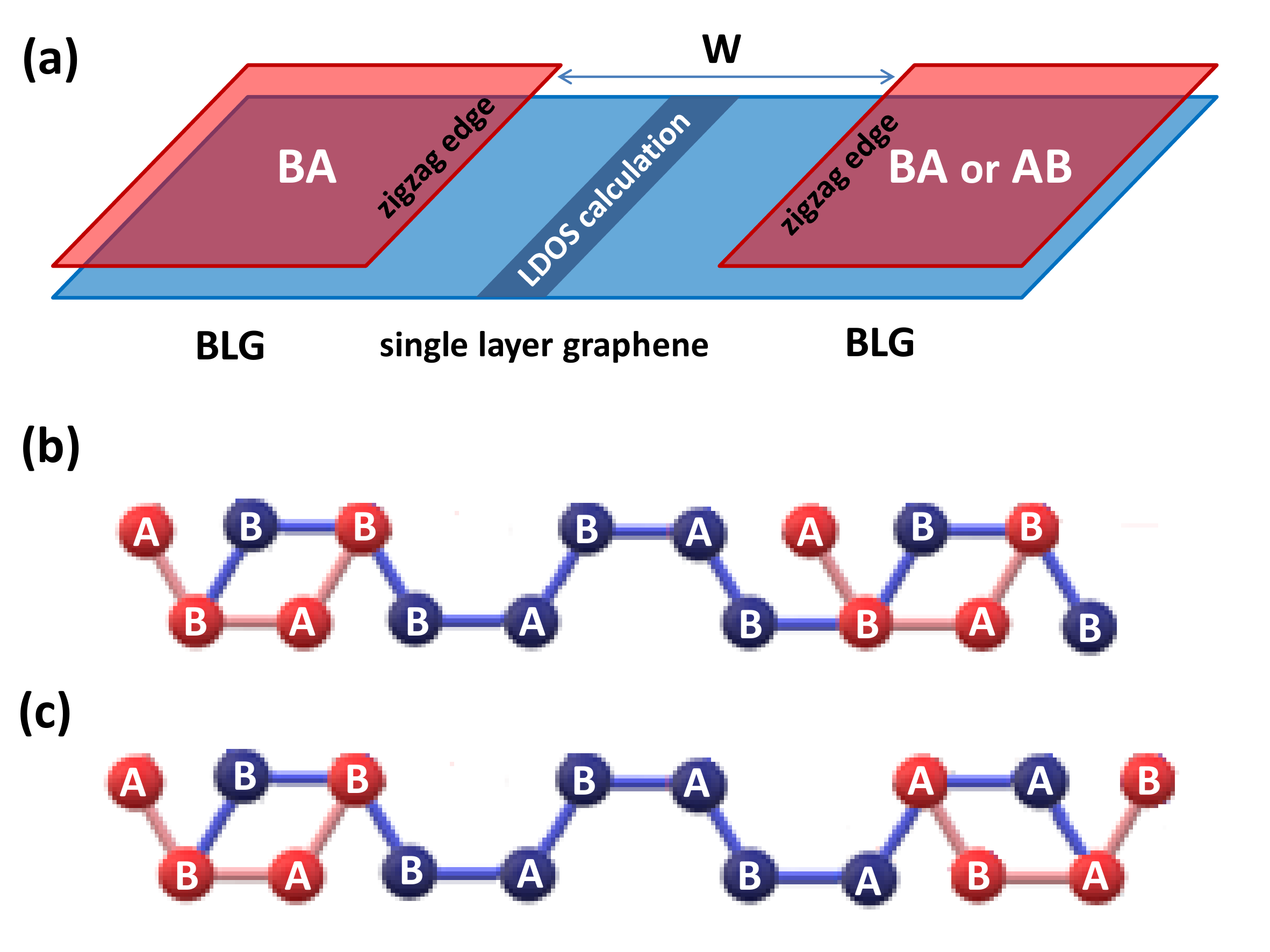}
\caption{\label{fig:first}
(a) Schematic representation of the investigated system. Two separated layers of graphene (red) are deposited on the single layer (blue), forming two bilayers connected by the single layer of width $W$. The BLGs may differ in the stacking order. The central strip of SLG (dark blue) of  width equal to one graphene unit cell, $W=1$, marks the area for the calculation of LDOS. Top view of the fragment structure with $W=1$ for two cases: (b) when both BLGs have the same stacking order of sublattices A and B, (c) when BLGs have different stacking orders. BLGs extend to infinity at the left and right sides; only one unit cell along the zigzag direction is presented.
}
 \end{figure}

In this paper we demonstrate that the energy spectrum characteristic for gated BLG, with its tunable energy gap, is induced in a SLG when the strip of the single layer is connected to two distant gated bilayers, as shown in Fig. \ref{fig:first} (a). The structure can also be seen as the SLG with additional layers deposited and forming two distant bilayers. Additionally, we show that when the stacking order in these bilayers is different, a topological gapless state connecting the induced valence and conduction band continua appears in the SLG. For some gate voltages and widths of the SLG slab, this state is localized mostly in the central part of the single layer. It means that one-dimensional current can flow in the single layer graphene. This finding can be exploited in graphene nanoelectronics and sheds new light on the nature of topological gapless states that so far were proved to exist only in BLG.
Topologically protected states appear in BLG at stacking boundaries, which are usually unintentional. Our finding suggest an easy way how to prepare it on demand, i.e., by proper deposition of two separated sheets of graphene on the SLG substrate.

\section{Model and methods}

The system we investigate consists of a large SLG (infinite for the purpose of calculation). On its the top  two half-infinite graphene layers are deposited at distance $W$ forming two bilayers connected by the SLG strip. The structure is schematically shown in Fig. \ref{fig:first} (a). We assume that the bilayers have Bernal stacking order of sublattices and are cut along the zigzag direction. 
Our aim is to study how the presence of bilayers influence the energy structure of the SLG slab. To do so we calculate the local density of states (LDOS) at the central part of the SLG. The LDOS depends on the wave vector $k$ due to the periodicity in the zigzag direction. All the calculations are performed using the Green function matching technique within the standard $\pi$-electron tight-binding approximation \cite{Datta,Chico_1996,Nardelli_1999,Jaskolski_2005}. Intra-layer and inter-layer hopping parameters $t=2.7$ eV and $\gamma_1 = 0.27$ eV are used, respectively \cite{Castro_2007,Ohta_2006}. We investigate LDOS for different widths $W$ measured in the numbers of graphene unit cells, and for different voltages $V$ applied to the bottom layer. Two values of $V$ are taken into account, $V < \gamma_1$ and $V > \gamma_1$, both accessible experimentally. This choice is dictated by the observation that topological gapless states in BLG with stacking domain walls localize in different layers depending on the value of $V$ vs $\gamma_1$ \cite{Jaskolski_2018}.

\begin{figure}[ht]
\centering
\includegraphics[width=\columnwidth]{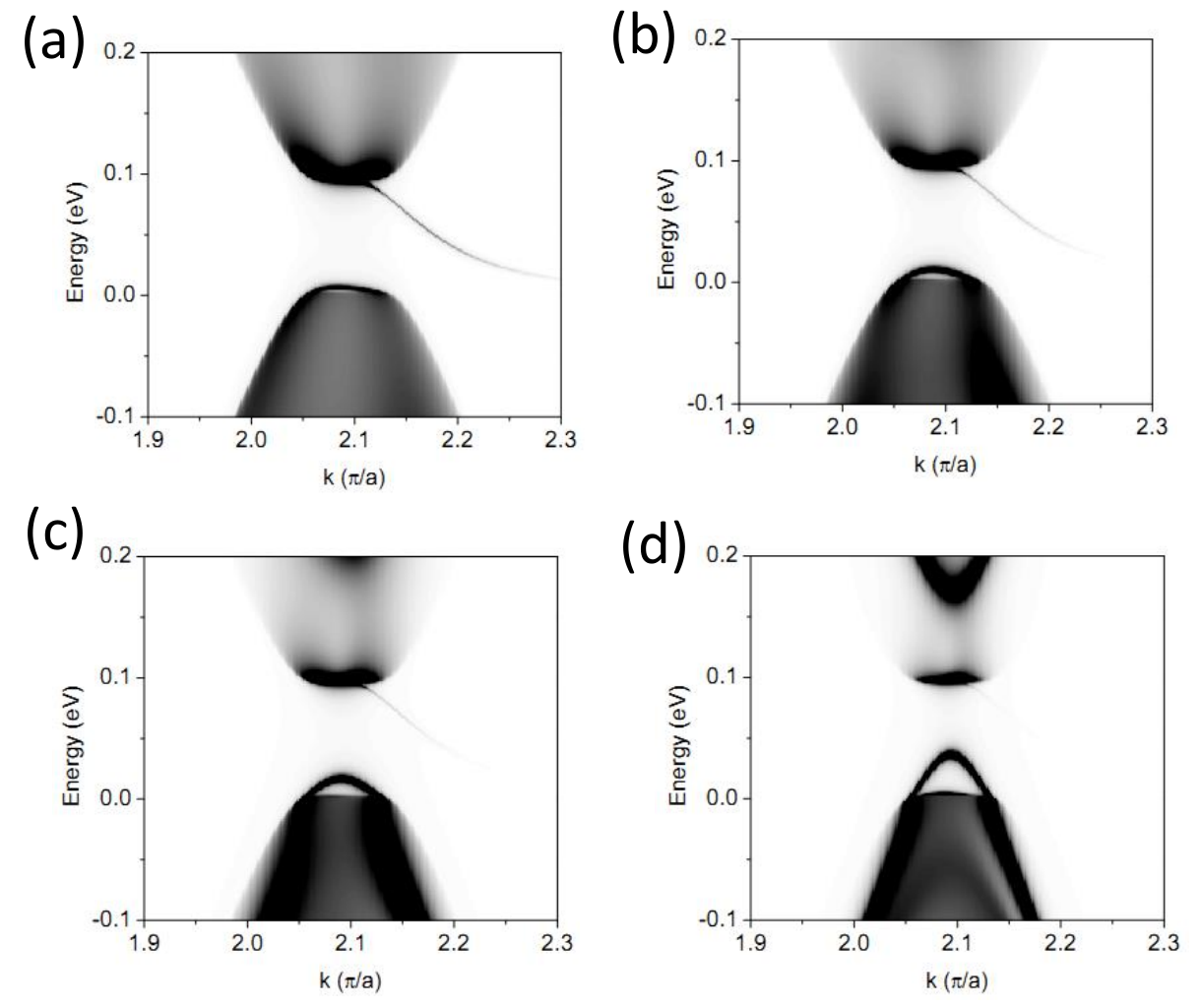}
\caption{\label{fig:second}
LDOS calculated in the central part of SLG (shown in Fig. \ref{fig:first}) for different widths $W$ of SLG. The voltage $V=0.1$ V is applied to the bottom layer. The BLGs have the same stacking order AB. (a)-(d) correspond to width $W=5$, 13, 21, and 42 graphene unit cells, respectively.
}
 \end{figure}

\section{Results}
We start with the calculation of LDOS in the central part of the SLG slab (see Fig. \ref{fig:first} (a)) for the gate $V=0.1$ eV.
We first consider the case when both BLGs have the same order of stacking sublattices, as visualized in Fig. \ref{fig:first} (b). The results for different widths $W$ of the SLG are shown in Fig. \ref{fig:second}. For $W=5$, 13, and 21 the LDOS does not differ substantially from the typical LDOS of the gated BLG. For $W$ larger than 5, a single band separates from the valence band (VB) continuum and starts to approach the conduction band (CB) edge. This is the state of the quantum well (QW) formed of the SLG slab embedded in between two BLG leads. For very large $W=42$, one can see another QW state approaching the cone from the conduction continuum region. We have checked that for growing $W$ more and more QW states appear in the spectrum, while the LDOS in the CB and VB continua weakens. The QW bands close the energy gap for $W \approx 50$, i.e., for $W \approx 20$ nm. It is important to note that the QW states are the only states that originate from the slab of the single layer; the CB and VB continua are induced in SLG by the nearby bilayers.

\begin{figure}[ht]
\centering
\includegraphics[width=\columnwidth]{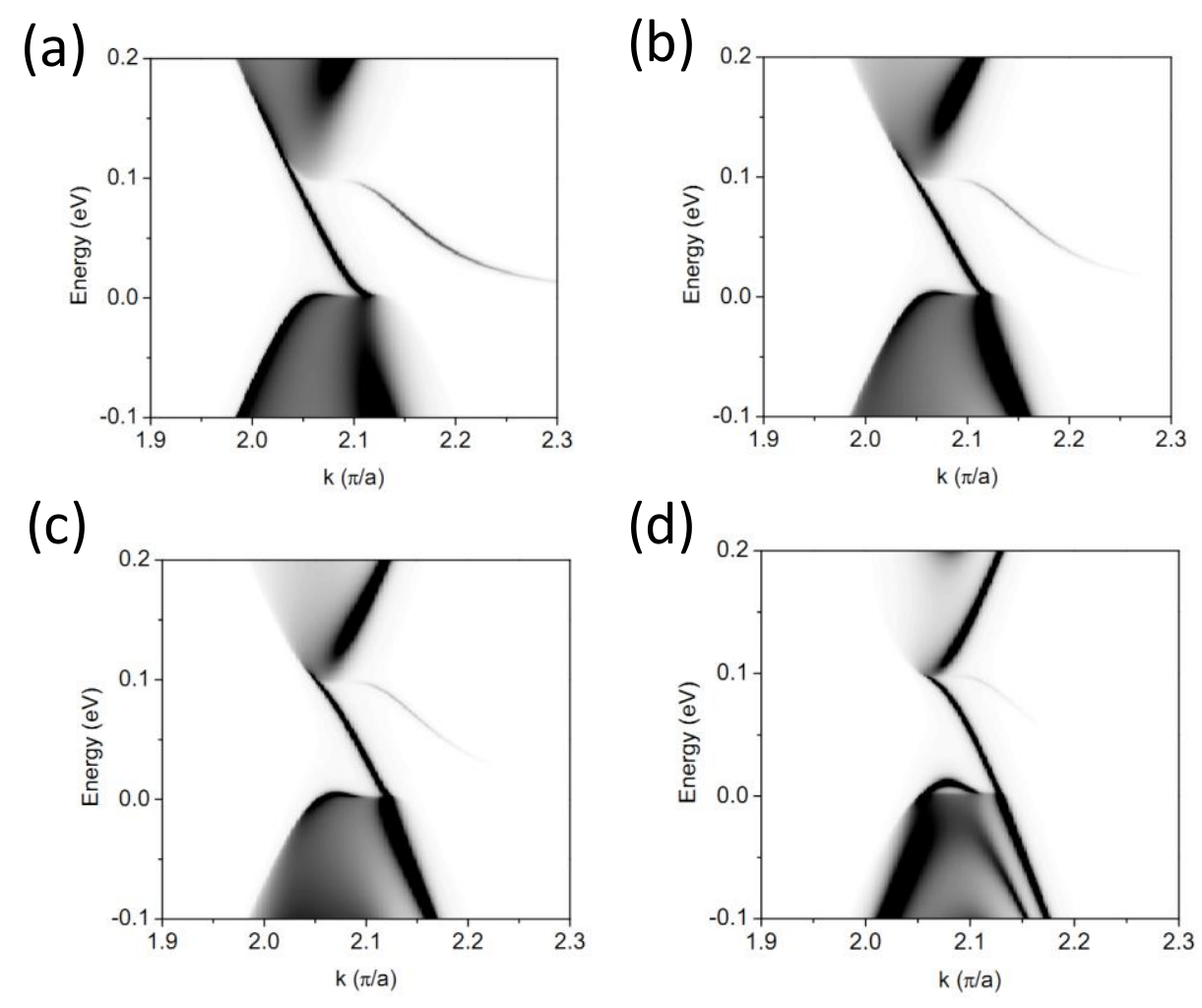}
\caption{\label{fig:third} 
LDOS calculated in the central part of SLG (shown in \ref{fig:first}) for different widths $W$ of SLG. The voltage $V=0.1$ V is applied to the bottom layer. The BLGs have different stacking orders AB and BA. (a)-(d) correspond to width $W=5$, 13, 21, and 42 graphene unit cells, respectively.
}
\end{figure}

A note is required about the thin LDOS structure seen in Fig. \ref{fig:second}, which is pinned to the conduction band at the cone and approaches the zero energy for larger values of $k$. This is the trace of the edge state localized at the zigzag-edge of the upper layer of the left BLG. The presence of such states has been reported also for SLG-BLG interfaces \cite{Peeters_2017}. As reported in \cite{Wakabayashi_2010,Jaskolski_2011_B83}, the zigzag-edge states of the SLG are localized at one sublattice only (red B sublattice in this case, see Fig. \ref{fig:first} (b)). However, this state couples via $\gamma_1$ to the bottom layer and thus its traces can be seen in the LDOS of the SLG slab. For $k=\pi /a$, the zigzag-edge state is localized exclusively at the outermost atoms of the edge, couples weakly to the bottom layer via one atom only (in the unit cell),  and is thus almost invisible in the calculated LDOS. At the cone, i.e., for $k=2/3 \pi /a$, the edge state decays slowly away from the edge, couples stronger to the bottom layer (via many atoms of the connected sublattices), and is thus well visible in the LDOS. The right BLG also has the zigzag edge. However, the corresponding edge state is localized at the nodes that are not connected to the bottom layer (read A sublattice in Fig. \ref{fig:first} (b)); this state is therefore invisible in the LDOS spectrum of the bottom layer.

\begin{figure}[ht]
\centering
\includegraphics[width=\columnwidth]{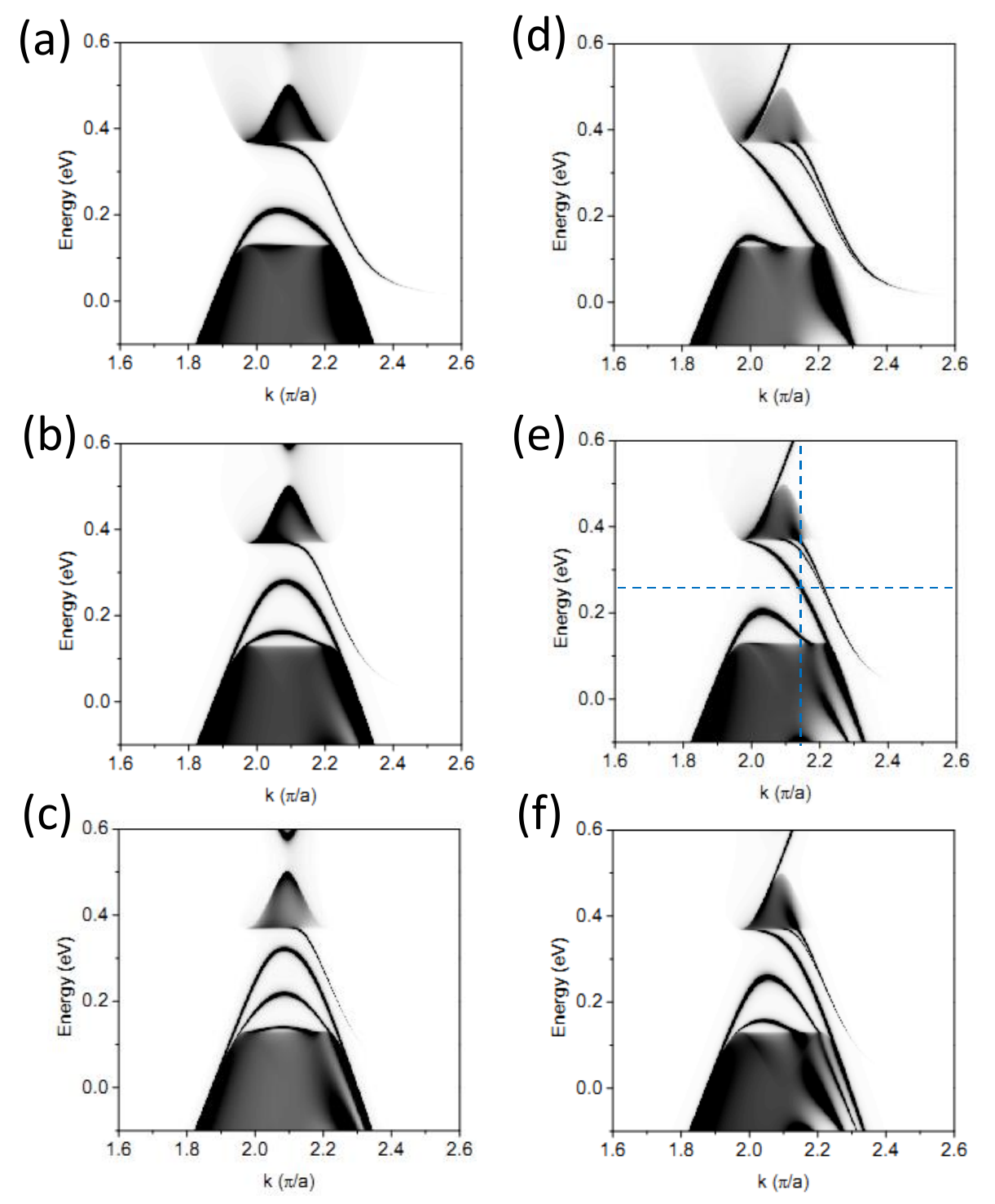}
\caption{\label{fig:fourth} 
LDOS calculated in the central part of SLG (shown in \ref{fig:first}) for different widths $W$ of the SLG. The voltage $V=0.5$ V is applied to the bottom layer. The BLGs have the same stacking order in (a)-(c), and different stacking order in (d)-(f). (a) and (d) correspond to width $W=5$, (b) and (e) to $W=13$, (c) and (f) to $W=21$ graphene unit cells.
The blue dashed lines in (e) show how the energy (in the middle of the gap) and the corresponding value of $k$ are selected for the calculation of the density distribution of the gapless states presented in Fig. \ref{fig:fifth}.
}
\end{figure}

When the BLGs have different order of stacking sublattices, a gapless state connecting the conduction and valence bad continua appears in the energy spectrum of the SLG slab. This is  shown in Fig. \ref{fig:third}. Unlike the case of typical stacking domain walls in gated BLG, where two topological states appear in the gap, here only one such state is present. As reported in \cite{Jaskolski_2018}, the gapless states induced in BLG by stacking domain walls are localized in different layers for $V \ne \gamma_1$. In the system studied in this report no upper layer is present in the region when the stacking order changes from BA to AB.  Therefore, only the gapless state localized in the bottom layer persists. 

Fig. \ref{fig:fourth} shows the LDOS calculated in the center of the SLG slab for a gate voltage of $V=0.5$ eV. The calculations have been performed for $W=5$, 13, and 21. The results for the case when both BLGs have the same stacking order are presented in the left column of the figure. The results for different stacking orders in BLGs are shown in the right column. The presented results are similar to those obtained for $V=0.1$ eV. Also for this gate voltage, the fundamental features of the gated BLG, like the valence and conduction band continua, the energy gap, and topological gapless state, are induced in the single layer. More QW states are visible in the gap for this gate voltage because the gap is situated deeper in the valence energy region of the bottom (gated) layer. 

\begin{figure}[ht]
\centering
\includegraphics[width=\columnwidth]{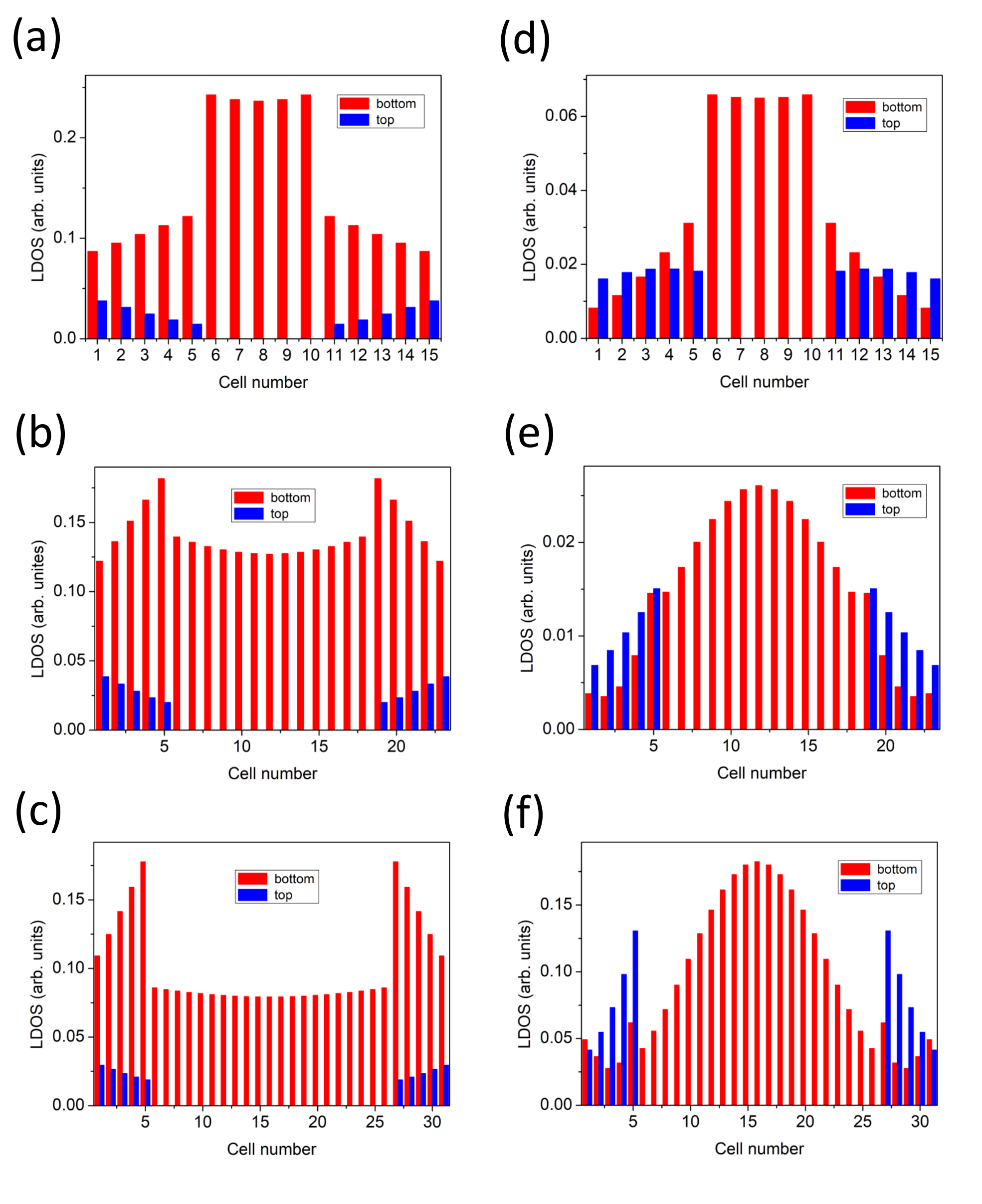}
\caption{\label{fig:fifth} 
Density distribution of topological gapless state for different width $W$ of the SLG. 
(a) and (d) correspond to width $W=5$, (b) and (e) to $W=13$, (c) and (f) to $W=21$ graphene unit cells. Left column - the results for gate $V=0.1$ V, right column – $V=0.5$ V. The LDOS distribution is calculated 
 for energy in the middle of the gap, equal $E=0.05$ eV for gate $V=0.1$ V, and $E=0.25$ eV for gate $V=0.5$ V. One vertical bar represents LDOS calculated for one four-atom unit cell. Five unit cells  are taken into account in the left and right BLG leads, so the SLG part starts at cell number 6.
LDOS distribution in the bottom and top layers are represented by red and blue bars, respectively.
}
\end{figure}

When the BLGs differ in the stacking order, the two edge states localized at the zigzag edges of the upper layers are connected to the bottom layer, see Fig. \ref{fig:first} (c). Therefore, they interact, mix,  and split, as seen in Fig. \ref{fig:fourth} (d)-(f). The splitting is present also in the case of $V=0.1$ eV, albeit much smaller and not visible within the resolution of LDOS presented in Fig. \ref{fig:third}.

We now investigate the density distribution of the gapless states. To this end,  we choose the energy in the middle of the gap and find the $k$ value  corresponding to a given gapless state, as shown, for example, in Fig. \ref{fig:fourth} (e). Next, for that particular pair ($E,k$), we calculate the layer resolved LDOS in each four-atom cell (ABAB) across the SLG, see Fig. \ref{fig:first} (b) and (c). Five four-atom cells in both layers of the BLGs are also taken into account. The results are presented in Fig. \ref{fig:fifth}.  For both values of $V$ and small width $W=5$ of the SLG,  the density of the gapless state is largely concentrated in the SLG slab. For $V=0.1$ eV, and increasing $W$, the density in the SLG slab slowly decreases comparing to the LDOS in the lower layer of the BLG. For $V=0.5$ eV, the effect is just the opposite; for increasing $W$, the density is stronger localized in the central part of the SLG. 

This behavior can be easily explained. When $W$ increases, more and more QW states appear in the energy gap and they push the topological state to the right (i.e., to larger values of $k$). As reported in \cite{Jaskolski_2018} for the case of typical stacking domain walls in BLG with two topological states in the energy gap, the gapless state that appears at the right hand side of the cone localizes in the upper layer when $V < \gamma_1$. However, this layer is absent in the system we investigate. Thus, for $V=0.1$, the density of the gapless state has to decrease in the SLG slab when its width $W$ increases. In contrast, the gapless state is pushed to the right with the increasing $W$ when $V > \gamma_1$, and thus has to stronger concentrate in the lower layer, i.e., in the SLG slab. This is our most important result; for $V > \gamma_1$ and large widths of the SLG, the gapless state strongly localizes in the central part of the SLG. 
This means that if the single layer of graphene is appropriately prepared, one can thus achieve strictly one-dimensional topologically protected currents for the use in nanoelectronic applicaations \cite{notka_tranz}.

\section{Conclusions}
We have investigated the electronic structure of a single layer graphene connected to a pair of gated bilayers. We have demonstrated that the energy structure characteristic for gated bilayer is induced to the single layer. Thus, we have shown how to force the single layer graphene to open the energy gap between the valence and conduction band continua. Moreover, we have demonstrated that when the bilayers differ in the stacking order of the sublattices, a topologically protected gapless state, appropriate for bilayers with stacking domain walls, appears in the gap and localizes in the single layer. For some values of the applied gate voltage and some separation distances between the bilayers, this gapless state is strongly confined to the central part of the SLG. Since the gapless states provide one-dimensional topologically protected currents, this finding can be potentially exploited in graphene nanoelectronics. 
Stacking domain walls with localized gapless states occur unintentionally in BLG. This work suggest also an easy way for experimental preparation of a bilayer system with a topologically protected state and associated one-dimensional current.



\end{document}